\documentclass{article}

\usepackage[psamsfonts]{amssymb}
\usepackage{amsmath}
\usepackage{cite}
\usepackage{bm}
\usepackage{graphicx}
\usepackage{yhmath}
\usepackage{bbm}

\def\d{\mathrm{d}}

\def\id{\mathbf{1}}

\def\ir{\mathrm{i}}

\begin{document}
\begin{titlepage}
\noindent{\large\textbf{Poisson structure on a space with linear
SU(2) fuzziness}}

\vskip 1 cm

\begin{center}
{Mohammad~Khorrami~{\footnote {mamwad@mailaps.org}}\\
\vskip 0.1cm
Amir~H.~Fatollahi~{\footnote {ahfatol@gmail.com}}\\
\vskip 0.1cm
Ahmad~Shariati~{\footnote {shariati@mailaps.org}}}
\vskip 10 mm
\textit{ Department of Physics, Alzahra University, Tehran
1993891167, Iran. }

\end{center}

\vspace{0.5cm}

\begin{abstract}
\noindent The Poisson structure is constructed for a model in
which spatial coordinates of configuration space are
noncommutative and satisfy the commutation relations of a Lie
algebra. The case is specialized to that of the group SU(2), for
which the counterpart of the angular momentum, as well as the
Euler parameterization of the phase space are introduced.
SU(2)-invariant classical systems are discussed, and it is
observed that the path of particle can be obtained by the solution
of a first-order equation, as the case with such models on
commutative spaces. The examples of free particle,
rotationally-invariant potentials, and specially the isotropic
harmonic oscillator are investigated in more detail.
\end{abstract}
\end{titlepage}

\section{Introduction}
There have been arguments supporting the idea that the ordinary
picture of spacetime breaks down at spacetime intervals in which
the quantum gravity effects would be important, the so-called
Planck length and time \cite{doplicher}. The reasonings lie on the
expectation that gravity, as the theory responsible for partial
properties of spacetime, would present an undetermined character
at scales where quantum effects are dominant. As a consequence,
one is inclined to believe in some kinds of space-space and
space-time uncertainty relations \cite{doplicher}, whose
appearances usually point to noncommutative objects. In fact, by a
reverse way of reasoning, it has been argued that one might had to
expect that a proper description of quantum gravity should be
possible based on noncommutative theories of spacetime
\cite{madore1,rivelles,yang,stein}. In the simplest case of
canonical noncommutative space the coordinates satisfy
\begin{equation}\label{kfs.1}
[\hat{x}_\mu,\hat{x}_\nu]=\ir\,\theta_{\mu\,\nu}\,\id,
\end{equation}
in which $\theta$ is an antisymmetric constant tensor and $\id$
represents the unit operator. It has been understood that the
longitudinal directions of D-branes in the presence of a constant
B-field background appear to be noncommutative, as seen by the
ends of open strings \cite{9908142,99-2,99-3,99-4}. The
theoretical and phenomenological implications of such
noncommutative coordinates have been extensively studied
\cite{reviewnc}. In particular, it is argued that the spacetime
background that emerges from the gravity theory based on canonical
noncommutativity corresponds to a flat one
\cite{madore1,rivelles,yang,stein}.

One direction to extend studies on noncommutative spaces is to
consider spaces where the commutators of the coordinates are not
constants. Examples of this kind are the noncommutative cylinder
and the $q$-deformed plane \cite{chai}, the so-called
$\kappa$-Poincar\'{e} algebra \cite{majid,ruegg,amelino,kappa},
and linear noncommutativity of the Lie algebra type
\cite{wess,sasak}. In the latter the dimensionless spatial
positions operators satisfy the commutation relations of a Lie
algebra:
\begin{equation}\label{kfs.2}
[\hat{x}_a,\hat{x}_b]= f^c{}_{a\, b}\,\hat{x}_c,
\end{equation}
where $f^c{}_{a\,b}$'s are structure constants of a Lie algebra.
One example of this kind is the algebra SO(3), or SU(2). A special
case of this is the so called fuzzy sphere \cite{madore,presnaj},
where an irreducible representation of the position operators is
used which makes the Casimir of the algebra,
$(\hat{x}_1)^2+(\hat{x}_2)^2+(\hat{x}_3)^2$, a multiple of the
identity operator (a constant, hence the name sphere). One can
consider the square root of this Casimir as the radius of the
fuzzy sphere. This is, however, a noncommutative version of a
two-dimensional space (sphere).

In \cite{0612013,fakE1,fakE2} a model was introduced in which the
representation was not restricted to an irreducible one, instead
the whole group was employed. In particular the regular
representation of the group was considered, which contains all
representations. As a consequence in such models one is dealing
with the whole space, rather than a sub-space, like the case of
fuzzy sphere as a 2-dimensional surface. In \cite{0612013} basic
ingredients for calculus on a linear fuzzy space, as well as basic
notions for a field theory on such a space, were introduced. In
\cite{fakE1} basic elements for calculating the matrix elements
corresponding to transition between initial and final states were
discussed. Models based on the regular representation of SU(2)
were treated in more detail, giving explicit forms of the tools
and notions introduced in their general forms
\cite{0612013,fakE1}. In \cite{fakE1} and \cite{fakE2} the tree
and 1-loop diagrams for a self-interacting scalar field theory
were discussed, respectively. It is observed that models based on
Lie algebra type noncommutativity enjoy three features:
\begin{itemize}
\item They are free from any ultraviolet divergences if the group
is compact. \item There is no momentum conservation in such
theories. \item In the transition amplitudes only the so-called
planar graphs contribute.
\end{itemize}
The reason for latter is that the non-planar graphs are
proportional to $\delta$-distributions whose dimensions are less
than their analogues coming from the planar sector, and so their
contributions vanish in the infinite-volume limit usually taken in
transition amplitudes \cite{fakE2}.

The facts that in such theories the mass-shell condition is
different, and there is no momentum conservation, lead to
different consequences (with respect to ordinary theories) in
collisions. This was exploited in \cite{skf}, where it was seen
that there may be a new threshold for the collision of two
massless particles to produce massive particles.

The purpose of the present work is to examine the calssical
mechanics defined on space with SU(2) fuzziness. In particular,
the Poisson structure induced by noncommutativity of SU(2) type is
investigated, as well as the consequences of rotational symmetry
in such spaces.

The scheme of the rest of this paper is the following. In section
2, the phase space corresponding to a space with Lie algebra type
noncommutativity is studied. Specifically, the corresponding
Poisson structure is investigated. In section 3, these are
specialized to the group SU(2). In section 4 classical systems are
studied which are SU(2)-invariant, and a formulation is presented
to obtain the path of the particle. Section 5 is devoted to some
examples.

\section{The Poisson structure}
Consider a Lie group G. Denote the members of a basis for the
left-invariant vector fields corresponding to this group by
$\hat{x}_a$'s. These fields satisfy (\ref{kfs.2}), with the
structure constants of the Lie algebra corresponding to G. The
coordinates $\hat{k}^a$ are defined such that
\begin{equation}\label{kfs.3}
U(\hat{\mathbf{k}}):=[\,\exp(\hat{k}^a\,\hat{x}_a)]\,U(\mathbf{0}),
\end{equation}
where $U(\hat{\mathbf{k}})$ is the group element corresponding to
the coordinates $\hat{\mathbf{k}}$, $U(\mathbf{0})$ is the
identity, and $\exp(\hat{x})$ is the flux corresponding to the
vector field $\hat{x}$. The action of $L_{\hat{x}_a}$ (the Lie
derivative corresponding to the vector field $\hat{x}_a$) on an
arbitrary scalar function $F$ can be written like
\begin{equation}\label{kfs.4}
L_{\hat{x}_a}(F)=\hat{x}_a{}^b\,\frac{\partial\,F}{\partial\hat{k}^b},
\end{equation}
where $\hat{x}_a{}^b$'s are scalar functions, and satisfy
\begin{equation}\label{kfs.5}
\hat{x}_a{}^b(\mathbf{k}=\mathbf{0})=\delta_a^b.
\end{equation}
One can define the vector fields $\hat{X}_a$ locally through
\begin{equation}\label{kfs.6}
L_{\hat{X}_a}(F)=\frac{\partial\,F}{\partial\hat{k}^a},
\end{equation}
so that
\begin{equation}\label{kfs.7}
\hat{x}_a=\hat{x}_a{}^b\,\hat{X}^b.
\end{equation}
Then, considering scalar functions as operators acting on scalar
functions through simple multiplications, and vector fields as
operators acting on scalar functions through Lie derivation, one
arrives at the following commutation relations
\begin{align}\label{kfs.8}
[\hat{X}_a,\hat{X}_b]&=0,\\ \label{kfs.9}
[\hat{X}_a,\hat{k}^b]&=\delta_a^b,\\ \label{kfs.10}
[\hat{k}^a,\hat{k}^b]&=0.
\end{align}
One should, however, remember that the functions $\hat{k}^a$ and
the vector fields $\hat{X}_a$ are only locally defined. One can
write the above commutation relations in terms of $\hat{x}_a$'s
instead of $\hat{X}_a$'s. The equation corresponding to
(\ref{kfs.8}) would be (\ref{kfs.2}), while that corresponding to
(\ref{kfs.9}) would be
\begin{equation}\label{kfs.11}
[\hat{x}_a,\hat{k}^b]=\hat{x}_a{}^b,
\end{equation}
and as $\hat{x}_a{}^b$'s are scalar functions, they commute with
$\hat{k}^a$'s.

Next consider the right-invariant vector fields
$\hat{x}_a^{\mathrm{R}}$, so that they coincide with their
left-invariant analogues at the identity of the group:
\begin{equation}\label{kfs.12}
\hat{x}_a^{\mathrm{R}}(\hat{\mathbf{k}}=\mathbf{0})=
\hat{x}_a(\hat{\mathbf{k}}=\mathbf{0}).
\end{equation}
These field satisfy the commutation relations
\begin{align}\label{kfs.13}
[\hat{x}_a^{\mathrm{R}},\hat{x}_b^{\mathrm{R}}]&=
-f^c{}_{a\,b}\,\hat{x}_c^{\mathrm{R}},\\ \label{kfs.14}
[\hat{x}_a^{\mathrm{R}},\hat{x}_b]&=0.
\end{align}
Using these, one defines the new vector field $\hat{J}_a$ through
\begin{equation}\label{kfs.15}
\hat{J}_a:=\hat{x}_a-\hat{x}_a^{\mathrm{R}}.
\end{equation}
These are the generators of the adjoint action, and satisfy the
commutation relations
\begin{align}\label{kfs.16}
[\hat{J}_a,\hat{J}_b]&=f^c{}_{a\,b}\,\hat{J}_c,\\
\label{kfs.17} [\hat{J}_a,\hat{X}_b]&=f^c{}_{a\,b}\,\hat{X}_c,\\
\label{kfs.18} [\hat{J}_a,\hat{x}_b]&=f^c{}_{a\,b}\,\hat{x}_c,\\
\label{kfs.19} [\hat{k}^c\,\hat{J}_a]&=f^c{}_{a\,b}\,\hat{k}^b.
\end{align}
Equations (\ref{kfs.9}), (\ref{kfs.17}), and (\ref{kfs.19}) show
that
\begin{equation}\label{kfs.20}
\hat{J}_a=-f^c{}_{a\,b}\,\hat{k}^b\,\hat{X}_c.
\end{equation}

To construct the phase space, all that is needed is to transform
the commutation relations to Poisson brackets. This can be done
through the correspondence $[.\,,.]/(\ir\hbar)\to\{.\,,.\}$.
However, one should also take care of the dimension of the
quantities, and their reality. To do so, let us define the
following quantities.
\begin{align}\label{kfs.21}
p^a&:=(\hbar/\ell)\,\hat{k}^a,\\
\label{kfs.22} X_a&:=\ir\,\ell\,\hat{X}_a,\\
\label{kfs.23} x_a&:=\ir\,\ell\,\hat{x}_a,\\
\label{kfs.24} x_a{}^b(\mathbf{p})&:=
\hat{x}_a{}^b[(\ell/\hbar)\,\mathbf{p}],\\
\label{kfs.25} J_a&:=\ir\,\hbar\,\hat{J}_a,
\end{align}
where $\ell$ is a constant of dimension length. One then arrives
at the following Poisson brackets.
\begin{align}\label{kfs.26}
\{p^a,p^b\}&=0,\\
\label{kfs.27} \{X_a,p^b\}&=\delta_a^b,\\
\label{kfs.28} \{X_a,X_b\}&=0,\\
\label{kfs.29} \{x_a,p^b\}&=x_a{}^b,\\
\label{kfs.30} \{x_a,x_b\}&=\lambda\,f^c{}_{a\,b}\,x_c,\\
\label{kfs.31} \{J_a,X_b\}&=f^c{}_{a\,b}\,X_c,\\
\label{kfs.32} \{J_a,x_b\}&=f^c{}_{a\,b}\,x_c,\\
\label{kfs.33} \{p^c,J_a\}&=f^c{}_{a\,b}\,p^b,\\
\label{kfs.34} \{J_a,J_b\}&=f^c{}_{a\,b}\,J_c,
\end{align}
where the dimension of $\lambda$ is that of inverse momentum:
\begin{equation}\label{kfs.35}
\lambda:=\frac{\ell}{\hbar}.
\end{equation}
Using (\ref{kfs.5}) it is seen that in the limit $\lambda\to 0$
(corresponding to $\ell\to 0$), the ordinary Poisson brackets are
retrieved.

\section{The group SU(2), and the Euler parameters}
For the group SU(2), one also can define the Euler parameters
through
\begin{equation}\label{kfs.36}
[\exp(\phi\,T_3)]\,[\exp(\theta\,T_2)]\,[\exp(\psi\,T_3)]:=
[\exp(k^a\,T_a)],
\end{equation}
where $T_a$'a are the generators of SU(2) satisfying the
commutation relation
\begin{equation}\label{kfs.37}
[T_a,T_b]=\epsilon^c{}_{a\,b}\,T_c.
\end{equation}
Using these, one arrives at
\begin{align}\label{kfs.38}
L_{\hat x_1}(F)&=-\frac{\cos\psi}{\sin\theta}\,\frac{\partial
F}{\partial\phi}+\sin\psi\,\frac{\partial F}{\partial\theta}+
\frac{\cos\psi\,\cos\theta}{\sin\theta}\,\frac{\partial
F}{\partial\psi},\\ \label{kfs.39} L_{\hat
x_2}(F)&=\frac{\sin\psi}{\sin\theta}\,\frac{\partial
F}{\partial\phi}+\cos\psi\,\frac{\partial F}{\partial\theta}-
\frac{\sin\psi\,\cos\theta}{\sin\theta}\,\frac{\partial
F}{\partial\psi},\\ \label{kfs.40} L_{\hat x_3}(F)&=\frac{\partial
F}{\partial\psi},
\end{align}
and
\begin{align}\label{kfs.41}
L_{\hat
J_1}(F)=\;&\frac{\cos\phi\,\cos\theta-\cos\psi}{\sin\theta}\,\frac{\partial
F}{\partial\phi}+(\sin\phi+\sin\psi)\,\frac{\partial
F}{\partial\theta}\cr &+
\frac{-\cos\phi+\cos\psi\,\cos\theta}{\sin\theta}\,\frac{\partial
F}{\partial\psi},\\ \label{kfs.42} L_{\hat
J_2}(F)=\;&\frac{\sin\phi\,\cos\theta+\sin\psi}{\sin\theta}\,\frac{\partial
F}{\partial\phi}+(-\cos\phi+\cos\psi)\,\frac{\partial
F}{\partial\theta}\cr & +
\frac{-\sin\phi-\sin\psi\,\cos\theta}{\sin\theta}\,\frac{\partial
F}{\partial\psi},\\ \label{kfs.43} L_{\hat
J_3}(F)=\;&-\frac{\partial F}{\partial\phi}+\frac{\partial
F}{\partial\psi},
\end{align}
for an arbitrary scalar field $F$. Again, using the
dimensionalization process of the previous section one arrives at
\begin{align}\label{kfs.44}
x_1=\;&\lambda\,\left[-\frac{\cos\psi}{\sin\theta}\,X_\phi+\sin\psi\,X_\theta+
\frac{\cos\psi\,\cos\theta}{\sin\theta}\,X_\psi\right],\\
\label{kfs.45}
x_2=\;&\lambda\,\left[\frac{\sin\psi}{\sin\theta}\,X_\phi+\cos\psi\,X_\theta-
\frac{\sin\psi\,\cos\theta}{\sin\theta}\,X_\psi\right],\\
\label{kfs.46} x_3=\;&\lambda\,X_\psi,\\
\label{kfs.47}
J_1=\;&\frac{\cos\phi\,\cos\theta-\cos\psi}{\sin\theta}\,X_\phi+
(\sin\phi+\sin\psi)\,X_\theta \cr &+
\frac{-\cos\phi+\cos\psi\,\cos\theta}{\sin\theta}\,X_\psi,\\
\label{kfs.48}
J_2=\;&\frac{\sin\phi\,\cos\theta+\sin\psi}{\sin\theta}\,X_\phi
+(-\cos\phi+\cos\psi)\,X_\theta\cr &+
\frac{-\sin\phi-\sin\psi\,\cos\theta}{\sin\theta}\,X_\psi,\\
\label{kfs.49} J_3=\;&-X_\phi+X_\psi,
\end{align}
where $\phi$, $\theta$, and $\psi$ are the canonical momenta
corresponding to the canonical coordinates $X_\phi$, $X_\theta$,
and $X_\psi$, respectively. $\phi$, $\theta$, and $\psi$ are
dimensionless, while $X_\phi$, $X_\theta$, and $X_\psi$ have the
dimension of action.

One also has
\begin{equation}\label{kfs.50}
\cos\frac{\hat k}{2}=\cos\frac{\theta}{2}\,
\cos\frac{\phi+\psi}{2},
\end{equation}
or in terms of the dimensionful quantities
\begin{equation}\label{kfs.51}
\cos\frac{ \lambda\,p}{2}=\cos\frac{\theta}{2}\,
\cos\frac{\phi+\psi}{2},
\end{equation}
where
\begin{align}\label{kfs.52}
\hat{k}&:=(\delta_{a\,b}\,\hat{k}^a\,\hat{k}^b)^{1/2},\nonumber\\
p&:=(\delta_{a\,b}\,p^a\,p^b)^{1/2}.
\end{align}

One could obtain $\hat x_a$'s in a different way, trying to use
only (\ref{kfs.2}) and (\ref{kfs.5}). It turns out that these are
not sufficient to determine $\hat x_a$'s uniquely \cite{sasak}.
However, defining $\hat x_a$'s as the left invariant vector fields
of the group manifold determines them uniquely. It can be shown
that adding
\begin{equation}\label{kfs.53}
[\hat{x}_a, \hat{k}]=\frac{\hat{k}_a}{\hat{k}}
\end{equation}
to (\ref{kfs.2}) and (\ref{kfs.5}), completely determines $\hat
x_a$'s. One then arrives at
\begin{equation}\label{kfs.54}
\hat{x}_a{}^b=\frac{\hat{k}}{2}\,\cot\frac{\hat{k}}{2}\,\delta_a^b
+ \left(1-\frac{\hat{k}}{2}\,\cot\frac{\hat{k}}{2}\right)\,
\frac{\hat{k}_a \,\hat{k}^b}{\hat{k}^2}+\frac{1}{2}\,
\epsilon_a{}^{b\,c}\,\hat{k}_c,
\end{equation}
which is equivalent to (\ref{kfs.38})-(\ref{kfs.40}).

\section{SU(2)-invariant classical systems}
Consider a configuration space with linear SU(2)-fuzziness and its
corresponding phase space. This is like what introduced in section
2, with $f$ equal to $\epsilon$. A classical system which is
characterized by a Hamiltonian $H$, is said to be SU(2)-invariant,
if $H$ is SU(2)-invariant, that is if the Poisson brackets of $H$
with $J_a$'s vanish. A Hamiltonian which is a function of only
$(\mathbf{p}\cdot\mathbf{p})$ and $(\mathbf{x}\cdot\mathbf{x})$ is
clearly so, where
\begin{equation}\label{kfs.55}
\mathbf{A}\cdot\mathbf{B}:=\delta_{a\,b}\,A^a\,B^b.
\end{equation}
From now on, assume that the system is so, that is the Hamiltonian
is a function of only $(\mathbf{p}\cdot\mathbf{p})$ and
$(\mathbf{x}\cdot\mathbf{x})$. Then $\mathbf{J}$ is a constant
vector and one can choose the axes so that the third axis is
parallel to this vector:
\begin{align}\label{kfs.56}
J_1&=0,\nonumber \\
J_2&=0.
\end{align}
Solving these, one arrives at
\begin{align}\label{kfs.57}
\phi+\psi&=0,\nonumber\\
X_\phi+X_\psi&=0.
\end{align}
Defining
\begin{align}\label{kfs.58}
\frac{\phi-\psi}{2}&=:\chi,\nonumber\\
X_\phi-X_\psi &=J,
\end{align}
it is seen by (\ref{kfs.49}) that
\begin{equation}\label{kfs.59}
\{J,\chi\}=1,
\end{equation}
and that $J$ and $\chi$ are in involution with $X_\theta$ and
$\theta$, so that $(X_\theta, J; \theta, \chi)$ are canonical
coordinates left after applying (\ref{kfs.57}). Applying
(\ref{kfs.57}), one arrives at
\begin{align}\label{kfs.60}
x_1&=\lambda\,\left(-\frac{J}{2}\,\frac{1+\cos\theta}{\sin\theta}\,\cos\chi-
X_\theta\,\sin\chi\right),\nonumber\\
x_2&=\lambda\,\left(-\frac{J}{2}\,\frac{1+\cos\theta}{\sin\theta}\,\sin\chi+
X_\theta\,\cos\chi\right),\nonumber\\
x_3&=\lambda\,\left(-\frac{J}{2}\right),\nonumber\\
\mathbf{x}\cdot\mathbf{x}&=\lambda^2\,\left[X_\theta^2+
\frac{J^2}{4}\,\left(1+\cot^2\frac{\theta}{2}\right)\right],\nonumber\\
\cos\frac{\hat{k}}{2}&=\cos\frac{\theta}{2}.
\end{align}
It is seen that the motion is not in the plane $x_3=0$, but in a
plane parallel to that, as $x_3$ does not vanish but is a
constant.

Defining $\rho$ and $\alpha$ (the so called polar coordinates
corresponding to $x_1$ and $x_2$) as
\begin{align}\label{kfs.61}
x_1&=:\rho\,\cos\alpha,\nonumber\\
x_2&=:\rho\,\sin\alpha,
\end{align}
one has
\begin{align}\label{kfs.62}
\mathbf{x}\cdot\mathbf{x}&=\rho^2+\frac{\lambda^2\,J^2}{4},\nonumber\\
\rho^2&=\lambda^2\,(X_\theta^2+J^2\,u^2),\nonumber\\
\alpha&=\chi-\tan^{-1}\frac{X_\theta}{J\,u},
\end{align}
where
\begin{equation}\label{kfs.63}
u:=\frac{1}{2}\,\cot\frac{\theta}{2}.
\end{equation}
One then has
\begin{equation}\label{kfs.64}
\mathbf{x}\cdot\mathbf{x}=\lambda^2\,\left[X_\theta^2+
J^2\,\left(\frac{1}{4}+u^2\right)\right].
\end{equation}
The Hamiltonian is a function of only $u$ and
$(\mathbf{x}\cdot\mathbf{x})$, where $(\mathbf{x}\cdot\mathbf{x})$
itself contains $J$ (a constant of motion), $u$, and $X_\theta$.
Our aim is now to reduce the problem to that of a system of one
degree of freedom. The aim is to write an equation for the path,
in terms of $\rho$ and $\alpha$. To do so, one notices by
(\ref{kfs.64}) that
\begin{align}\label{kfs.65}
\frac{\d u}{\d t}&=\{u, H\},\nonumber\\
&=\lambda^2\,\frac{\partial
H}{\partial(\mathbf{x}\cdot\mathbf{x})}\,(2\,X_\theta)\,
\left(\frac{1}{4}+u^2\right),
\end{align}
and
\begin{align}\label{kfs.66}
\frac{\d\chi}{\d t}&=\{\chi, H\},\nonumber\\
&=-\lambda^2\,\frac{\partial
H}{\partial(\mathbf{x}\cdot\mathbf{x})}\,(2\,J)
\,\left(\frac{1}{4}+u^2\right).
\end{align}
Using these, one arrives at
\begin{equation}\label{kfs.67}
X_\theta=-J\,\frac{\d u}{\d\chi},
\end{equation}
so that
\begin{align}\label{kfs.68}
\rho^2&=\lambda^2\,J^2\,\left[\left(\frac{\d
u}{\d\chi}\right)^2+u^2\right],\\ \label{kfs.69}
\alpha&=\chi+\tan^{-1}\left(\frac{1}{u}\,\frac{\d
u}{\d\chi}\right).
\end{align}
Using these, one arrives at
\begin{align}\label{kfs.70}
\frac{\d\rho}{\d\chi}&=\lambda^2\,\frac{J^2}{\rho}\, \frac{\d
u}{\d\chi}\,
\left(u+\frac{\d^2 u}{\d\chi^2}\right)\nonumber\\
\frac{\d\alpha}{\d\chi}&=u\,\left(u+\frac{\d^2
u}{\d\chi^2}\right)\, \left[u^2+\left(\frac{\d
u}{\d\chi}\right)^2\right]^{-1},\nonumber\\
&=\lambda^2\,\frac{J^2}{\rho^2}\,u\,\left(u+\frac{\d^2
u}{\d\chi^2}\right),
\end{align}
from which
\begin{equation}\label{kfs.71}
\frac{\d\rho}{\d\alpha}=\frac{\rho}{u}\, \frac{\d u}{\d\chi}.
\end{equation}
Using this and (\ref{kfs.68}), one can eliminate $(\d u/\d\chi)$
and arrive at
\begin{equation}\label{kfs.72}
\frac{1}{u^2}=\lambda^2\,J^2\,\left[\frac{1}{\rho^2}+\frac{1}{\rho^4}\,
\left(\frac{\d\rho}{\d\alpha}\right)^2\right].
\end{equation}
The Hamiltonian is a function of $\hat k$ and
$(\mathbf{x}\cdot\mathbf{x})$. One can express it in terms of $u$
and $(\mathbf{x}\cdot\mathbf{x})$, and arrive at
\begin{equation}\label{kfs.73}
H\left\{\frac{1}{u^2}=\lambda^2\,J^2\,\left[\frac{1}{\rho^2}+
\frac{1}{\rho^4}\, \left(\frac{\d\rho}{\d\alpha}\right)^2\right],
\mathbf{x}\cdot\mathbf{x}=\rho^2+\frac{\lambda^2\,J^2}{4}\right\}=E,
\end{equation}
where $E$ is the energy. This is a first order differential
equation for the path of the system, which should be compared to
the corresponding commutative case ($\lambda\to 0$):
\begin{equation}\label{kfs.74}
H\left\{\mathbf{p}\cdot\mathbf{p}=J^2\,\left[\frac{1}{\rho^2}+
\frac{1}{\rho^4}\, \left(\frac{\d\rho}{\d\alpha}\right)^2\right],
\mathbf{x}\cdot\mathbf{x}=\rho^2\right\}=E,
\end{equation}
where it is understood that in the commutative limit
$(\lambda\,u)$ has been kept constant (equal to the inverse of the
momentum).

A special situation is when the angular momentum vanishes. In this
case the differential equation (\ref{kfs.73}) is not a good
equation, as $J$ vanishes and $(\d\rho/\d\alpha)$ goes to
infinity. Using (\ref{kfs.66}) it is seen that $\chi$ is constant,
from which and (\ref{kfs.69}) it turns out that $\alpha$ is
constant as well. So $\phi$ and $\psi$ are constants and $X_\phi$
and $X_\psi$ vanish. One then arrives at an effectively
one-degree-of-freedom Hamiltonian (involving $\theta$ and
$X_\theta$), from which a relation between $\theta$ and $X_\theta$
is found to be
\begin{equation}\label{kfs.75}
H(u, \mathbf{x}\cdot\mathbf{x}=\lambda^2\,X_\theta^2)=E.
\end{equation}
One can also write a first-order differential equation for
$\theta$, using (\ref{kfs.65}). To do so, one obtains $X_\theta$
from (\ref{kfs.65}), and inserts it in (\ref{kfs.75}).

\section{Examples}
Here we consider a case where the Hamiltonian is the sum of a
Kinetic term, which is a function of only $p$, and a potential
term, which is a function of only $(\mathbf{x}\cdot\mathbf{x})$.
Following \cite{0612013,fakE1,fakE2}, the kinetic term is taken to
be
\begin{equation}\label{kfs.76}
K=\frac{4}{\lambda^2\,m}\,\left(1-\cos\frac{\lambda\,p}{2}\right)
\end{equation}
for a particle of mass $m$. This function is increasing in $p$, as
long as $p$ does not exceed $(2\,\pi/\lambda)$, and periodic in
$p$ with the period $(4\,\pi/\lambda)$, showing that it is a
function of the group manifold SU(2). It is also reduced to
$p^2/(2\,m)$ for small values of $p$, which shows that for small
momenta the commutative results are obtained. One can express this
kinetic term in terms of $u$:
\begin{equation}\label{kfs.77}
K=\frac{4}{\lambda^2\,m}\,\left(1-\frac{2\,u}{\sqrt{1+4\,u^2}}\right).
\end{equation}

For a free particle, the Hamiltonian is equal to the kinetic term,
so that $u$ is a constant. Then using (\ref{kfs.72}) one arrives
at
\begin{equation}\label{kfs.78}
(\rho^{-1})^2+\left(\frac{\d\rho^{-1}}{\d\alpha}\right)^2=C^{-2},
\end{equation}
where $C$ is a constant. The solution to (\ref{kfs.79}) is
\begin{equation}\label{kfs.79}
\rho\,\cos(\alpha -\alpha_0)=C,
\end{equation}
where $\alpha_0$ is another constant. This, combined with the fact
that $x_3$ is a nonvanishing constant, describes a line in a plane
parallel to the plane $x_3=0$. For a commutative space, one would
obtain a line in the plane $x_3=0$.

If the particle is not free, one can still use (\ref{kfs.73}) to
obtain a differential equation for $\rho$, similar to the equation
of the commutative case. To do so, one obtains $u$ in terms of
$K$:
\begin{equation}\label{kfs.80}
\frac{1}{u^2}=4\,\left[\left(1-\frac{\lambda^2\,m}{4}\,K\right)^{-2}-1\right].
\end{equation}
Expressing the kinetic term in terms of the energy and the
potential term, and using (\ref{kfs.73}), one arrives at
\begin{equation}\label{kfs.81}
\frac{1}{\rho^2}+\frac{1}{\rho^4}\,\left(\frac{\d\rho}{\d\alpha}\right)^2=
\frac{4}{\lambda^2\,J^2}\,\left\{\left[1-\frac{\lambda^2\,m}{4}\,(E-V)\right]^{-2}
-1\right\},
\end{equation}
where $V$ is the potential energy which is a function of only
$(\mathbf{x}\cdot\mathbf{x})$. Also note that
$(\mathbf{x}\cdot\mathbf{x})$ is not equal to $\rho^2$, but is
obtained from (\ref{kfs.62}).

A special case is when the angular momentum vanishes. Then, using
(\ref{kfs.76}) one arrives at
\begin{equation}\label{kfs.82}
H=\frac{4}{\lambda^2\,m}\,\left(1-\cos\frac{\lambda\,p}{2}\right)+
V(\mathbf{x}\cdot\mathbf{x}=x^2),
\end{equation}
where
\begin{equation}\label{kfs.83}
x:=\lambda\,X_\theta,
\end{equation}
showing that
\begin{equation}\label{kfs.84}
\{x,p\}=1.
\end{equation}
Equations (\ref{kfs.82}) and (\ref{kfs.84}) describe a
one-degree-of-freedom system, the only difference of which with
the commutative case is in the kinetic term. Of course it is also
known that the coordinate $p$ is a periodic one, with the period
$(4\,\pi/\lambda)$.

An example is the case of a simple harmonic potential:
\begin{equation}\label{kfs.85}
V=\frac{1}{2}\,m\,\omega^2\,x^2,
\end{equation}
where $\omega$ is a constant. One then has
\begin{equation}\label{kfs.86}
H=\frac{4}{\lambda^2\,m}\,\left(1-\cos\frac{\lambda\,p}{2}\right)+
\frac{1}{2}\,m\,\omega^2\,x^2,
\end{equation}
which is like the Hamiltonian of a simple pendulum, with the roles
of $x$ and $p$ interchanged.

 \vspace{.5cm}

\noindent\textbf{Acknowledgement}:  This work was partially
supported by the research council of the Alzahra University.

\end{document}